\documentstyle[twocolumn,prb,aps]{revtex}
\tighten

\input epsf 

\begin{document}

\preprint{}
\draft

\title{Fermion-boson transmutation and comparison of statistical
ensembles in one dimension}

\author{K.\ Sch\"onhammer and V.\ Meden}
\address{Institut f\"ur Theoretische Physik der Universit\"at
G\"ottingen, Bunsenstr.\ 9, D-37073 G\"ottingen, Germany}

\date{9 August 1995}

\maketitle

%\vspace{-1.1cm}
%\widetext
\begin{abstract}

%\begin{center}
%\begin{minipage}{14.0cm}   
The theoretical description of
interacting fermions in one spatial dimension is simplified by the
fact that the low energy excitations can be described in terms of
bosonic degrees of freedom. This fermion-boson transmutation (FBT)
which lies at the heart of the Luttinger liquid concept is 
presented in a way which does not require a knowledge of quantum
field theoretical methods. As the basic facts can already be
introduced for noninteracting fermions they are mainly discussed.
As an application we use the FBT to present
exact results for the low temperature thermodynamics and the 
occupation numbers in the microcanonical and the canonical ensemble.
They are compared with the standard grand canonical results.  
%\end{minipage}
%\end{center}
\end{abstract}
%\vspace{-0.6cm}
%\begin{center}
%\begin{minipage}{14.5cm}
%\pacs{}
%
%
%\end{minipage}
%\end{center}
%]   
%\twocolumn

%
%\narrowtext    

\section{Introduction}

The low temperature thermodynamic properties of simple metals can
be qualitatively understood in terms of the simple Sommerfeld model
\cite{AM}, which treats the conduction electrons as noninteracting
fermions in a box. Typical results are a specific heat {\it linear}
in temperature and a constant spin susceptibility in qualitative
agreement with experiments. Landau's {\it Fermi liquid} theory
\cite{Landau} rests on the assumption of
{\it quasi-particles} which are in a one-to-one
correspondence to noninteracting fermions. This leads 
to a linear specific heat and a constant spin susceptibility but
involves {\it renormalized} quantities like the effective mass
and the quasi-particle interaction parameters \cite{Landau,NP}, which are 
difficult to calculate microscopically.
The consistency of the approach was shown using perturbation
theory to infinite order and more recently by
renormalization group techniques \cite{Shankar}.

The problem of interacting                    
fermions simplifies in {\ one dimension}. In a pioneering paper
\cite{Tomonaga} Tomonaga treated the case of a two-body interaction
which is long ranged in real space. He showed that the low energy
excitations of the noninteracting as well as the interacting system
can be described in terms of {\it noninteracting bosons} \cite{reprints}. The
important idea to solve the case of interacting fermions
was the observation that a long range interaction in real
space is short ranged in momentum space and therefore only particles and
holes in the vicinity of the Fermi points are involved in the 
interacting ground state and states with a low excitation energy.
To obtain his results, Tomonaga {\it linearized} the energy dispersion
around the two Fermi points $\pm k_F$.
Luttinger \cite{Luttinger} later used a model with strictly 
linear energy dispersions which is closely related to the massless
Thirring model \cite{Thirring}. The exact solution for the Luttinger
model was presented by Mattis and Lieb \cite{ML}. A very elegant
method to calculate correlation functions 
for the model is the bosonization of the 
fermion field operator \cite {LP,Haldane}.
The exponents of the anomalous power law decay of various correlation 
functions are determined by  the {\it anomalous dimension}, which can be 
calculated explicitly for the Tomonaga-Luttinger (TL-) model\cite
{Luttinger,ML,LP}.
It was an important
observation of Haldane\cite{Haldane} that the low energy physics
of the exactly solvable TL-model provides the
{\it generic} scenario for one-dimensional fermions with repulsive
interactions. Like in the Landau Fermi liquid picture \cite{Landau}
a few parameters completely determine the low energy physics. 
Generally they are as difficult to calculate as the Landau
parameters. In contrast to the higher dimensional case there are
additional exactly solvable models for which Haldane's {\it Luttinger
liquid} scenario can be tested and the parameters determining the
anomalous dimension can be calculated using the Bethe-ansatz
technique\cite{Schulz}. 

Another important manifestation of Luttinger liquids is 
called {\it spin charge separation}, i.e. 
for low energy excitations the charge and spin degrees
are completely decoupled. This shows up, for example, in the spectral
function of the one-particle Green's function \cite{MS,Voit} which
largely determines the photoemission spectrum. Recent high resolution
photoemission experiments on quasi one-dimensional organic conductors 
have been interpreted to show Luttinger liquid behaviour \cite{Baer}.

In the present paper we present the basic ideas behind the
Luttinger liquid concept in very simple terms, 
involving only {\it basic} quantum mechanics and statistical
mechanics. We therefore believe that our approach requires
less theoretical knowledge than the field theory approach to
FBT published in this journal some time ago \cite{Galic}. 

In section II we present the fermion-boson transmutation in terms that should be
accessible to anybody who has had an elementary course in solid state physics.
Especially {\it no} use is made of the method of second quantization. 
The concepts introduced in section II are used in section III and IV to compare 
the different statistical ensembles for noninteracting fermions in a box 
\cite{Reif,terHaar}. 
In the grand canonical ensemble 
the constraint of a fixed particle number which makes the calculation
complicated in the canonical ensemble is relaxed and the
determination
of thermodynamical quantities is straightforward. The microcanonical
ensemble is seldom used in this context. Our discussion of the
basic ideas of the FBT allows us to present a comparison 
of the three ensembles. We show explicitly {\it how} the results
approach each other in the limit of infinite box size with constant
density, i.e. the thermodynamic limit. For the calculation of the occupation
numbers in the canonical ensemble the numerical effort increases quickly with 
the system size when the elementary approach of section IV is used. We take this 
as an opportunity to discuss the fermion-boson transmutation again on a
more advanced level using the method of second quantization in section V.
The techniques which allow the calculation of the properties of the TL-model 
of {\it interacting} fermions \cite{Tomonaga,Luttinger} are introduced again
in the context of noninteracting fermions. The power of this method is shown 
by presenting a closed expression for the occupation number of noninteracting
fermions in the canonical ensemble, which allows to discuss the transition 
to the Fermi function in the thermodynamic limit.

The paper is written such that readers not familiar with the method of second
quantization can understand fermion boson transmutation in its elementary form
{\it and} the comparison of the statistical ensembles by reading sections I to IV
only.

\section{noninteracting fermions in a box}

We consider particles in one dimension in a box of length $L$ with
fixed boundary conditions $\phi(0)=\phi(L)=0$. The single particle
eigenfunctions and energies are given by

\begin{equation}\label{eq:1}
\phi_n(x)=\sqrt{2/L}\sin{(k_nx)},
\end{equation}
and
\begin{equation}\label{equ2}
\epsilon_n\equiv \epsilon(k_n)=\frac{\hbar^2k_n^2}{2m}   ,
\end{equation}
with $k_n=n\pi/L$ and $n\in {\rm I \! N}$. Later we will also consider periodic
boundary conditions. Then the spacing of the $k$-values is doubled and 
negative {\it and} positive values occur. In the following the spin of the
fermions is neglected for simplicity, i.e. we consider {\it spinless}
fermions as in [\onlinecite{Tomonaga,Luttinger}]. 
The ground state of $N$ fermions is given by filling the lowest
momentum states up to the Fermi wavevector $k_F=n_F\pi/L$  where
$n_F=N$ and the excited states can be classified by giving the
{\it occupation numbers} $n_{k_n}$, which are zero or one. For fixed
particle number $N$ the sum of the occupation numbers has to be equal
to $N$. This constraint makes the evaluation of the {\it canonical }
partition function difficult. In the well known procedure of Darwin and
Fowler \cite{terHaar} an integral representation for the Kronecker delta 
describing the constraint is used. If the
corresponding integral is evaluated by the method of steepest descent
the result is equivalent to the grand canonical calculation, where
only the average particle number is fixed as particle exchange
with a bath is assumed. In the following we show that in the {\it low 
temperature} regime the canonical partition function can be
expressed as the partition function of a photon
gas, i.e. massless bosons. 

We first summarize the well
known grand canonical results \cite{Reif}. The total energy $E_{gc}$ as a
function of the temperature $T$ and the chemical potential $\mu$
is given by
\begin{equation}\label{eq:3}
E_{gc}=\sum_n \frac{\epsilon_n}{e^{\beta [\epsilon_n-\mu]}+1} ,
\end{equation}
where $\beta=1/(k_BT)$ and $k_B$ is the Boltzmann constant.
The chemical potential $\mu$ becomes a function of temperature if
the average particle number is assumed to be given by the {\it fixed}
value $N$
\begin{equation}\label{eq:4}
N=\sum_n \frac{1}{e^{\beta[\epsilon_n-\mu(T)]}+1}. 
\end{equation}
In the thermodynamic limit $L\to \infty, N/L$ fixed, the low
temperature result for the energy, the specific heat and other
thermodynamic quantities can be explicitly evaluated using the 
Sommerfeld technique \cite{AM,Reif}. They are determined by the 
one-particle density of states per unit length $\rho_0(\epsilon)
=1/\pi \hbar v(\epsilon)$ in the {\it vicinity} of the Fermi energy $\epsilon_F
\equiv\epsilon_{n_F}$, where $\hbar$ times the velocity $v(\epsilon)$ is 
given by  $d\epsilon(k)/dk$ as a function of $\epsilon$.
The low temperature specific heat, for example,
is {\it linear} in temperature
and determined by $\rho(\epsilon_F)$ \cite {AM,Reif}.
This is an indication that in order to arrive at this result
one is allowed to {\it simplify} the spectrum of eigenvalues in Eq.\ (2)
in the vicinity of the Fermi energy.

For $n\gg 1$ the eigenvalue spectrum given by Eq.\ (2) becomes locally
{\it equidistant}
\begin{equation}\label{eq:5}
\frac{\epsilon_{n+1}-\epsilon_n}{\epsilon_n-\epsilon_{n-1}}
=\frac{2n+1}{2n-1}\to 1.
\end{equation}       
\begin{figure}[hbt]
\hspace{0.0cm} 
\epsfxsize8.0cm  %Breite des Plots (Siehe auch dvips)
\epsfbox{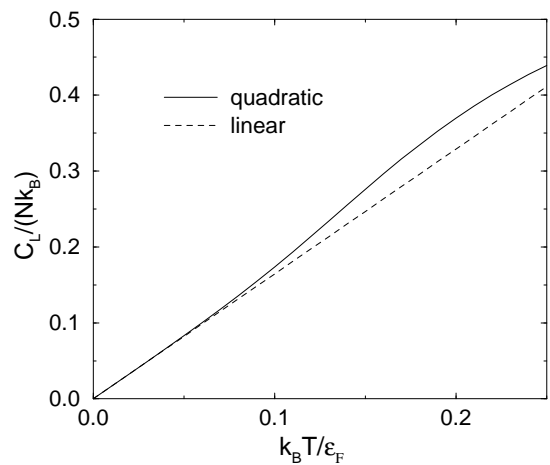}    %Evtl noch \label usw.
\vspace{0.0cm}
\caption{Specific heat for the quadratic (solid line) and the linearized
(dashed line) electron dispersions.}  
\end{figure}   
As in Refs. [\onlinecite{Tomonaga}] and [\onlinecite{Luttinger}] we therefore {\it linearize}
the spectrum $\epsilon_n$ around the Fermi point $n_F$
\begin{equation}\label{eq:6}
\epsilon_n^{(l)} =\epsilon_F +\hbar v_F(n-n_F)\pi/L,
\end{equation}                                       
where $v_F = \hbar k_F /m = \hbar \pi (N/L)/m$ is the Fermi velocity. As
we keep the density $(N/L)$ constant while increasing $L$ the Fermi velocity 
is {\it independent} of the system size.
In the following we consider this as the model for which we want to
determine the low temperature thermodynamics. Fig.\ 1 shows the 
temperature regime $k_BT \ll \epsilon_F$
for which the specific heats for both models agree.
The results where obtained by numerical evaluation of Eqs.\ (3)  and (4)
in the thermodynamic limit.  
The linearization in Eq.\ (6) is the essential step to obtain the exact
solution for the {\it interacting} TL-model \cite{Tomonaga,ML}.
The constant energy separation of the linearized model is given
by $\Delta\equiv \hbar v_F\pi/L$. 

For the linearized model also the excitation spectrum of the
$N$-electron system is equidistant
\begin{equation}\label{eq:7}
E_M=E_0 +M\Delta,
\end{equation}
where $E_0$ is the ground state energy of the filled Fermi sea.
For large $M$ the eigenvalues are {\it highly degenerate}. The 
determination of the degeneracy can be reduced to an old problem
in combinatorics \cite{Euler}, if we classify the corresponding
states not by fermionic occupation numbers as discussed above, but
in terms of the {\it upward shifts of the occupied levels
compared to the Fermi sea} \cite{reprints}. This 
is exemplified in Fig.\ 2 where a special excited state for $M=20$
is shown. The highest occupied level is shifted by seven energy
units,
the second and third by four, the forth by three and the fifth and
sixth highest by one unit. All deeper levels are unshifted. The
excitation energy in units of $\Delta$ is given by
$M=7+4+4+3+1+1=20$. This is one of the possible {\it partitions}
\cite {Euler,Andrews,SM} of the number $M=20$.
\begin{figure}[hbt]
\hspace{2.0cm} 
\epsfxsize4.0cm  %Breite des Plots (Siehe auch dvips)
\epsfbox{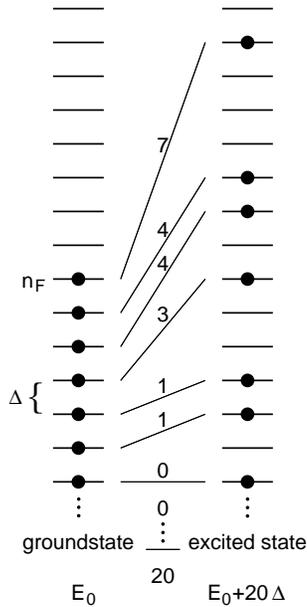}    %Evtl noch \label usw.
\vspace{0.0cm}
\caption{Example of the classification scheme for the excited states.}  
\end{figure}   
Each other of the 
$627$ partitions of
$M=20$, written in lexicographic order, as in our example,
uniquely classifies an other possible excited state
corresponding to this excitation energy.
For small $M$ it is easy to write down all partitions. For $M=4$
the five possible partitions are: $[4]; [3,1]; [2,2]; [2,1,1]; [1,1,1,1]$.
For a general $M$ a
partition is defined by a set of $M$ numbers $m_i\in {\rm I \! N}_0$ such that
\begin{equation}\label{eq:8}
M=m_1+2m_2+3m_3+... +Mm_M =\sum_{l=1}^M lm_l
\end{equation}
Already more than two hundred years ago
Euler \cite{Euler} proved important theorems about
the number of partitions.
His use of generating functions can be considered the earliest 
precursor of the methods of quantum statistical mechanics.

A formally simple expression for the number $g_M$ of partitions of $M$
can be given as a sum over a Kronecker delta using the defining 
Eq.\ (8)
\begin{equation}\label{eq:9}
g_M=\sum_{m_1=0}^{\infty}... \sum_{m_M=0}^{\infty}  
\sum_{m_{M+1}=0}^{\infty}... \;\; \delta_{M, \sum_{l=1}^{\infty}lm_l} .
\end{equation}
From Eq.\ (8) it is obvious that it is unnecessary to include more than 
$M$ summations in Eq.\ (9), but the form with the additional infinite 
number of summations
turns out to be useful to calculate the canonical partition
function of the linearized model
\begin{equation}\label{eq:10}
Z_c=\sum_{M=0}^{\infty}\tilde g_M e^{-\beta(E_0+M\Delta)} .
\end{equation} 
Here $\tilde g_M$ is the degeneracy corresponding to the excitation energy
$M\Delta$.  
For excitation energies $M \Delta$ smaller than $n_F \Delta$ all partitions
of $M$ lead to a possible excited state of the $N$-particle system and
$\tilde{g}_M = g_M$. For $M > n_F$ the degeneracy of excited states is smaller
than the number of partitions $\tilde{g}_M < g_M$. For $M=n_F+1$, e.g.\
there is no allowed excited $N$-particle state related to the partition 
\begin{eqnarray}
\underbrace{[1,1,...,1]}_{n_F+1},  \nonumber
\end{eqnarray}
because in the groundstate there are
only $n_F$ fermions that can be shifted upwards.  
For $M > n_F$ we have to take into account that 
one-particle states exist only for quantum numbers $n\geq 1$.
Two arguments can be given to use $g_M$ as the degeneracy also for $M>n_F$.
The one in spirit of
Tomonaga's\cite{Tomonaga} treatment of the interacting case is
to argue that for $n_F\Delta/(k_BT)\gg 1$ the contribution of these
highly excited states is {\it exponentially small} in this large 
ratio and the replacement of $\tilde{g}_M$ by $g_M$ in Eq.\ (10) makes 
no relevant difference.
The solution corresponding to Luttinger's \cite{Luttinger}
treatment of interacting fermions consists of {\it adding an 
 infinite Dirac sea} of fermions in one-particle states with
$n\leq 0 $. Then the replacement of $\tilde g_M $ by 
$g_M$ is exact, but one has to be careful with the treatment
of an infinite number of electrons \cite{Luttinger,ML}.  

After the replacement $\tilde g_M \to g_M$ the calculation of the
partition function $Z_c$ in Eq.\ (10) is simple if we use Eq.\ (9)
\begin{eqnarray}\label{eq:11}
Z_c & = & e^{-\beta E_0}\sum_{M=0}^{\infty}\sum_{m_1=0}^{\infty}
...\sum_{m_M=0}^{\infty}... \;\; \delta_{M,\sum_{l=1}^{\infty}lm_l} \nonumber \\*
&& \times  
e^{-\beta\Delta\sum_{l=1}^{\infty}lm_l}  \nonumber\\*
& = &e^{-\beta E_0}\prod_{l=1}^{\infty}\left( \sum_{m_l=0}^{\infty}e^{
-m_l\beta l\Delta} \right)      \nonumber\\*
& = & e^{-\beta E_0}\prod_{l=1}^{\infty}\frac{1}{1-e^{-\beta l\Delta}} .
\end{eqnarray}
This is the partition function of a system of massless bosons
or harmonic oscillators with energy $l\Delta= \hbar v_F k_l$. The $m_l$ which
describe the partitions play the roles of {\it bosonic quantum
numbers}.
{\it This is FBT in its most elementary form}.

If we differentiate $-\ln{Z_c}$ with respect to $\beta$
we obtain the canonical energy $E_c$
\begin{equation}\label{eq:12}
E_c(T)=E_0+\sum_{l=1}^{\infty}\frac{l\Delta}{e^{\beta l\Delta}-1}.     
\end{equation}
The evaluation of Eq.\ (12) for finite systems will be discussed in
the next section. Here we first want to show that the results for 
the energy per unit length $e(T)\equiv [E(T)-E_0]/L$
using Eq.\ (3) and Eq.\ (12) agree in the thermodynamic limit.
In this limit the sums can be replaced by integrals and we obtain
from Eq.\ (12)
\begin{equation}\label{eq:13}
e_c(T)=\frac{(k_BT)^2}{\pi\hbar v_F}I_{-} ,
\end{equation}
where the integrals $I_{\pm}$ are defined as
\begin{equation}\label{eq:14}
I_{\pm}\equiv \int_0^{\infty}\frac{x}{e^x\pm 1}dx .
\end{equation} 
Eq.\ (13) is the one-dimensional version of the Stefan-Boltzmann
law \cite{Reif} of black body radiation which states that 
in $d$ dimensions the energy density is proportional to $T^{d+1}$,
or to Debye's theory of the low temperature specific heat due
to acoustical phonons\cite{AM}. 

For the calculation in the grand canonical ensemble
using Eq.\ (3) the contribution
of the states above $\epsilon_F$ equals the part below $\epsilon_F$
for the linear dispersion and the chemical potential is
temperature {\it independent}. This yields

\begin{equation}\label{eq:15}
e_{gc}(T)=\frac{(k_BT)^2}{\pi\hbar v_F}2I_{+} .
\end{equation}
The integrals $I_{\pm}$ can be found in every good table of
integrals or textbook on statistical mechanics
\cite{Reif}. The fact that the results in Eqs.\ (13) and
(15) agree can be shown without actually calculating the integrals.
Elementary algebra yields
\begin{equation}\label{eq:16}
I_{-}-I_{+}=2\int_0^{\infty}\frac{x}{e^{2x}-1}dx =I_{-}/2.
\end{equation}
and we obtain $I_{-}=2I_{+}$, which shows the equivalence of the
canonical and the grand canonical ensemble in the thermodynamic
limit. The numerical value is given by \cite{Reif}
$I_{-}=\zeta (2)=\pi^2/6$ where $\zeta(x)$ is the Riemann zeta
function. The specific heat per unit length $c_L \equiv C_L/L$ for
both ensembles is therefore in the thermodynamic limit given by 
\begin{eqnarray}
\label{eq:16a}
c_{\infty}= \frac{\pi}{3} \frac{k_B^2 T}{\hbar v_F} .
\end{eqnarray}
In the following section the specific heat for both ensembles 
is discussed for finite systems. For a given density and temperature 
there is a natural length scale (thermal de Broglie wave length) given by             
$a(T) \equiv \hbar v_F \pi /(k_B T)$. The system size $L$ is large compared
to $a(T)$ if $k_B T$ is large compared to the level spacing $\Delta$.   

So far we have not discussed the microcanonical ensemble in which
one works with a fixed energy\cite{Reif} and the entropy as a 
function of energy is determined by the logarithm of the degeneracy.
Therefore we have for the linearized model
\begin{equation}\label{eq:17}
S_{micro}(E_M)=k_B \ln {(g_M)} ,
\end{equation}
where $g_M$ is the number of partitions. There are 
recursive methods to determine $g_M$ for small and intermediate
values of $M$ and excellent estimates can be given for large $M$
\cite{Andrews,Abramowitz}. 
We show also in the following section how the microcanonical results
approach those of the canonical and the grand canonical ensemble
for large $L$. 
  
\section{Thermodynamic properties for finite systems} 
The Stefan-Boltzmann law Eq.\ (13) for the energy per unit length is only correct 
in the thermodynamic limit. For finite $L$ the sum in Eq.\ (12) has to be performed 
numerically. For $L/a(T) \gg 1$ the Euler summation formula  in its most 
simple form \cite{Abramowitz}
\begin{equation}\label{eq:22}
\sum_{j=j_0}^{j=j_1}f(j)=\int_{j_0}^{j_1}f(x)dx +[f(j_0)+f(j_1)]/2
\mp...   ,
\end{equation}  
provides the leading finite size correction 
\begin{eqnarray}
\label{}
\frac{c_{L,c}}{c_{\infty}} = 1- \frac{3}{2 \pi^2} \frac{a(T)}{L}  
+ {\cal{O}} \left( \left[ \frac{a(T)}{L} \right]^2 \right) .
\end{eqnarray}
Fig.\ 3  shows $c_{L,c}/c_{\infty}$ as a function of $L/a(T)$. 
As $L/a(T) = k_B T/\Delta $ this plot can be viewed as the specific heat ratio
as a function of  system size for fixed temperature or as a function of
temperature for a given length of the system. For small arguments the ratio is 
small due to the well known phenomenon of freezing of the excited states for 
temperatures small compared to the excitation energies \cite{Reif}.
Fig.\ 3 also shows the corresponding results for the {\it grand
canonical} specific heat following from the numerical evaluation
of the result using Eq.\ (3). For small arguments the exponential
suppression is weaker than in the canonical case as the chemical
potential lies in the {\it middle} between the highest occupied and the
lowest unoccupied state of the Fermi sea.
The use of the Euler summation formula Eq.\ (19) shows that there is
{\it no} finite size correction of order $a(T)/L$ 
as in the canonical result Eq.\ (20) and   
the asymptotic limit is reached very quickly. 

\begin{figure}[hbt]
\hspace{0.0cm} 
\epsfxsize8.0cm  %Breite des Plots (Siehe auch dvips)
\epsfbox{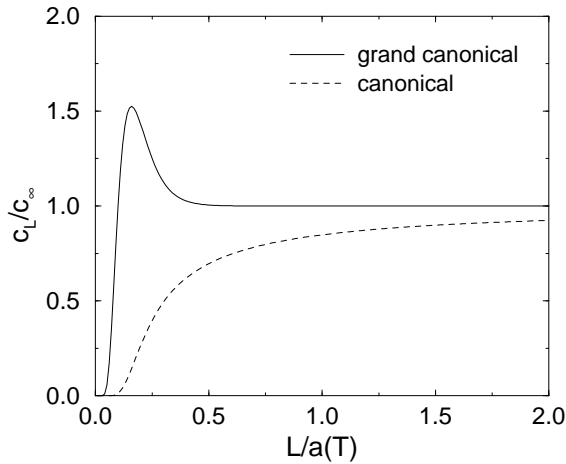}    %Evtl noch \label usw.
\vspace{0.0cm}
\caption{Ratio of the grand canonical specific heat and the specific heat 
in the thermodynamic limit $c_{\infty}$ (solid line) and   
the canonical specific heat and $c_{\infty}$ (dashed line) as a function
of $L/a(T)$.}  
\end{figure}   

As discussed in the preceding section the number of partitions
$g_M$ plays a central role for the thermodynamics of the linearized
model. In the microcanonical ensemble the number enters
explicitly in Eq.\ (18). The values of the $g_M$ for $M$ up to $500$ can be
found in a standard handbook  \cite{Abramowitz}. They increase very
quickly with $M$ as can be seen from $g_{500}\approx 2.3 \cdot 10^{21}$,
which leads to $S_{micro}(E_0+500\Delta)/k_B\approx 49.20$. For values
of $M$ of the order fifty or larger already the crudest
approximation of Ramanujan \cite{Andrews,Abramowitz}
\begin{equation}\label{eq:18}
(g_M)_R=e^{\pi \sqrt{2M/3}}/(4M\sqrt{3}),
\end{equation}
provides a good approximation. One obtains, for example
$\ln{ \left\{g_M(500)\right\}_R}\approx 49.18$.
In Fig.\ 4 we show $S_{micro}$ as a function of $M$ and compare
the exact result using the recursion relation\cite{Andrews,Abramowitz}
with the Ramanujan approximation Eq.\ (21), which smoothes the number 
theoretical fluctuations. The inverse microcanonical temperature
is defined as
\begin{equation}\label{eq:19}
\frac{1}{T_{micro}(E_M)}=\left[ \frac{\partial
S_{micro}(E_M)}{\partial E_M}\right]_{L} =
\frac{k_B}{\Delta}\left[\ln{g_M}\right]' .
\end{equation}
Due to the discreteness of the energy the derivative in Eq.\ (22) 
has to be evaluated as a ratio of differences 
$f_M'=(f_{M+1}-f_{M-1})/2$. This definition produces smoother
results than the definition involving differences of neighbouring
numbers.
The microcanonical definition of the heat capacity 
for constant $L$ is
\begin{equation}\label{eq:20}
C_{L,micro}(E_M)=\left[\left(\frac{\partial T_{micro}(E_M)}
{\partial E_M}\right)_L\right]^{-1}  .
\end{equation}
\begin{figure}[hbt]
\hspace{0.0cm} 
\epsfxsize8.0cm  %Breite des Plots (Siehe auch dvips)
\epsfbox{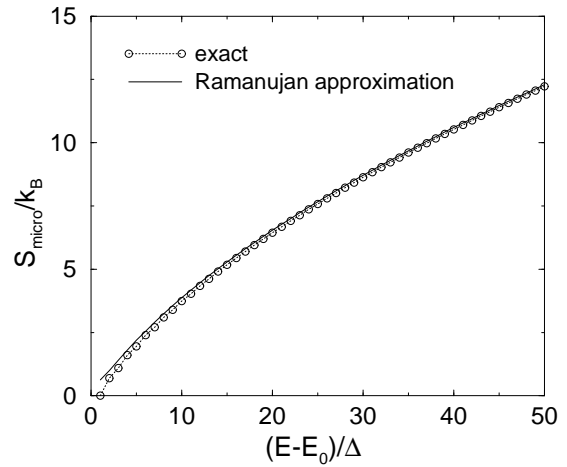}    %Evtl noch \label usw.
\vspace{0.0cm}
\caption{Microcanonical entropy as a function of the energy calculated with help
of the exact degeneracies (circles) and the Ramanujan approximation Eq.\ (18)
(solid line).}  
\end{figure}    
\begin{figure}[hbt]
\hspace{0.0cm} 
\epsfxsize8.0cm  %Breite des Plots (Siehe auch dvips)
\epsfbox{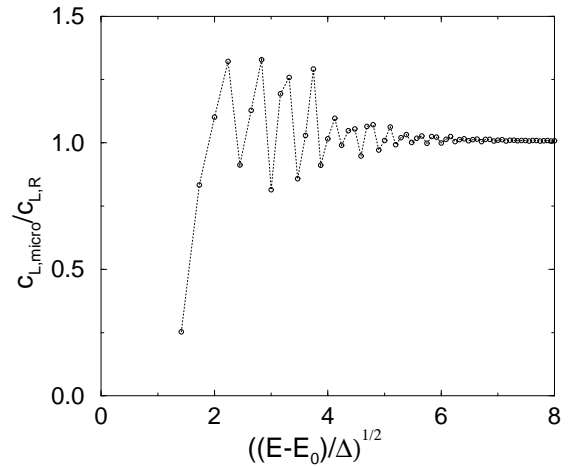}    %Evtl noch \label usw.
\vspace{0.0cm}
\caption{Ratio of the exact microcanonical specific heat and the one calculated 
with the help of the Ramanujan approximation as a function of 
$ \left[(E_M-E_0) / \Delta \right]^{1/2} $.}  
\end{figure}   
For $M \gg 1$ an analytical approximation for this quantity can be obtained using 
the Ramanujan expression Eq.\ (21), which we denote as $C_{L,R}$ 
\begin{eqnarray}
\label{}
C_{L,R}(E_M) & = & \pi k_B  \sqrt{\frac{2 (E_M-E_0)}{3 \Delta}} \nonumber \\
&&  + {\cal{O}} \left(
\sqrt{\frac{\Delta}{E_M-E_0}} \right) .
\end{eqnarray}
This should be compared with the relation $L c_{\infty} =
 \pi k_B \sqrt{2 L e(T) /(3 \Delta)}$, where $e(T) = e_c(T) = e_{gc}(T)$, which 
follows from Eqs.\ (13) or (15). As in the large energy limit $E-E_0$ in the 
microcanonical ensemble and $E(T)-E_0$ in the canonical (or grand canonical) 
ensemble can be identified, as discussed in the next section, $C_{L,R}/L$ agrees
with $c_{\infty}$ in this limit. Fig.\ 5 shows how the exact microcanonical 
result Eq.\ (23) approaches $C_{L,R}$ as a function of $\sqrt{(E_M-E_0)/\Delta}$,
which takes the role of $L/a(T)$ in Fig.\ 3. The ratio $c_{L,micro}/c_{L,R}$ shows
quite strong fluctuations for values $\sqrt{(E_M-E_0)/\Delta} < 5$. 

As in the thermodynamic limit 
the entropy per unit length agrees for all three ensembles we can
use the {\it grand canonical entropy} to obtain an estimate for the
{\it number of partitions}. This yields the Ramanujan expression Eq.\ (21)
with $4M\sqrt{3}$ replaced by $1$.

\section{Occupation numbers}
For variable particle number the method of second quantization
in Fock space is the appropriate mathematical framework.
The Hamiltonian for noninteracting fermions considered in
section II reads
\begin{equation}\label{eq:25}
H_0=\sum_m \epsilon_m\hat n_m=\sum_m\epsilon_m c_m^{\dagger}
c_m                                           ,
\end{equation}
where the $\hat n_m$ are the occupation number operators and
$c_m^{\dagger}$ $(c_m)$ are fermionic creation (annihilation)
operators, which obey canonical anticommutation relations
\begin{equation}\label{eq:26}
[c_m,c_l]_+=0 ,\;\; [c_m,c_l^{\dagger}]_+=\delta_{m,l}          .
\end{equation}
Explicit use of the method of second quantization is only made in the next
section. Here we discuss the calculation of the expectation value
of $\hat n_m$ in the three ensembles. The calculation in the 
grand canonical ensemble is standard \cite{Reif} and 
for arbitrary system size yields
the {\it Fermi function}
\begin{equation}\label{eq:27}
\left< \hat n_m \right >_{gc,T}=\frac{1}{e^{\beta (\epsilon_m-
\mu)}+1}\equiv f(\epsilon_m) .
\end{equation}

We next consider the model with the linearized dispersion Eq.\ (6) in 
the {\it microcanonical ensemble}. In section II
we have labeled the states of excitation energy $M\Delta$ in terms of
the {\it list of upward shifts} $[\alpha]_M$ 
\begin{equation}\label{eq:28}
[\alpha]_M =[l_1,l_2,...]  \;\;  \mbox{with} \;\; l_i \geq l_{i+1} \geq 0,
\end{equation}
which corresponds to a partition in lexicographic order.
The expectation value of the occupation number is given by
\begin{equation}\label{eq:29}
\left<\hat n_m \right>_{micro,E_M}=\frac{1}{g_M}
\sum_{[\alpha]_M} \left<[\alpha]_M \vert \hat
n_m \vert [\alpha]_M\right> .
\end{equation}
The expectation value of $\hat n_m$ in the state $\left\vert
[\alpha]_M\right >$
can easily be calculated due to the definition of the list $[\alpha]_M$
\begin{equation}\label{eq:30}
\left<[\alpha]_M\vert\hat n_m\vert [\alpha]_M\right>
=1   \;\;\; \mbox{for} \;\;
m=n_F+l_i-i+1,
\end{equation}
and zero otherwise. In order to perform the numerical calculations
all partitions of $M$ have to be {\it explicitly} created, which leads
to a drastic increase of computer time for increasing $M$. The
numerical results will be discussed below together with
the {\it canonical} result
\begin{equation}\label{31}
\left<\hat n_m\right>_{c,T}=\frac{1}{Z_c}\sum_{M=0}^{\infty}
g_M\left<\hat n_m\right>_{micro,E_M}e^{-\beta(E_0+M\Delta)} .
\end{equation}
The canonical result can therefore be obtained directly from
the microcanonical one if the temperature is low enough that
the probability distribution
\begin{equation}\label{eq:32}
w_M(T)=g_M e^{-\beta(E_0+M\Delta)}/Z_c,
\end{equation}
has decayed to ``zero'' for values of $M$ for which the calculation
of the microcanonical average is still feasible. Note that while the
Fermi function can be obtained from the grand canonical potential
by a derivative with respect to $\epsilon_m$  \cite{Reif} such a procedure is
{\it not} possible for the canonical free energy.
In order to calculate the canonical average for higher temperatures 
the method of the {\it bosonization of the field operator}\cite
{LP,Haldane} can be used which is described in the next section.
We therefore only consider results for quite low temperatures
in this section.

Fig.\ 6 shows the average occupation numbers for the three ensembles
for $\tau \equiv k_B t/\Delta =3$. 
The microcanonical curve is for $M=10$, which
roughly corresponds to the maximum of the 
probability distribution $w_M(T)$ shown in Fig.\ 7. The overall
agreement of the results is quite good already at this small value
of $\tau$. Note that the microcanonical result yields the
ground state occupancies for $\vert m-n_F\vert > M$, e.g.\ 
$\left<\hat n_m \right>_{micro,E_M}=0$ for $m > M+n_F$, while 
the canonical and the grand canonical results go to zero continuously.

\begin{figure}[hbt]
\hspace{0.0cm} 
\epsfxsize8.0cm  %Breite des Plots (Siehe auch dvips)
\epsfbox{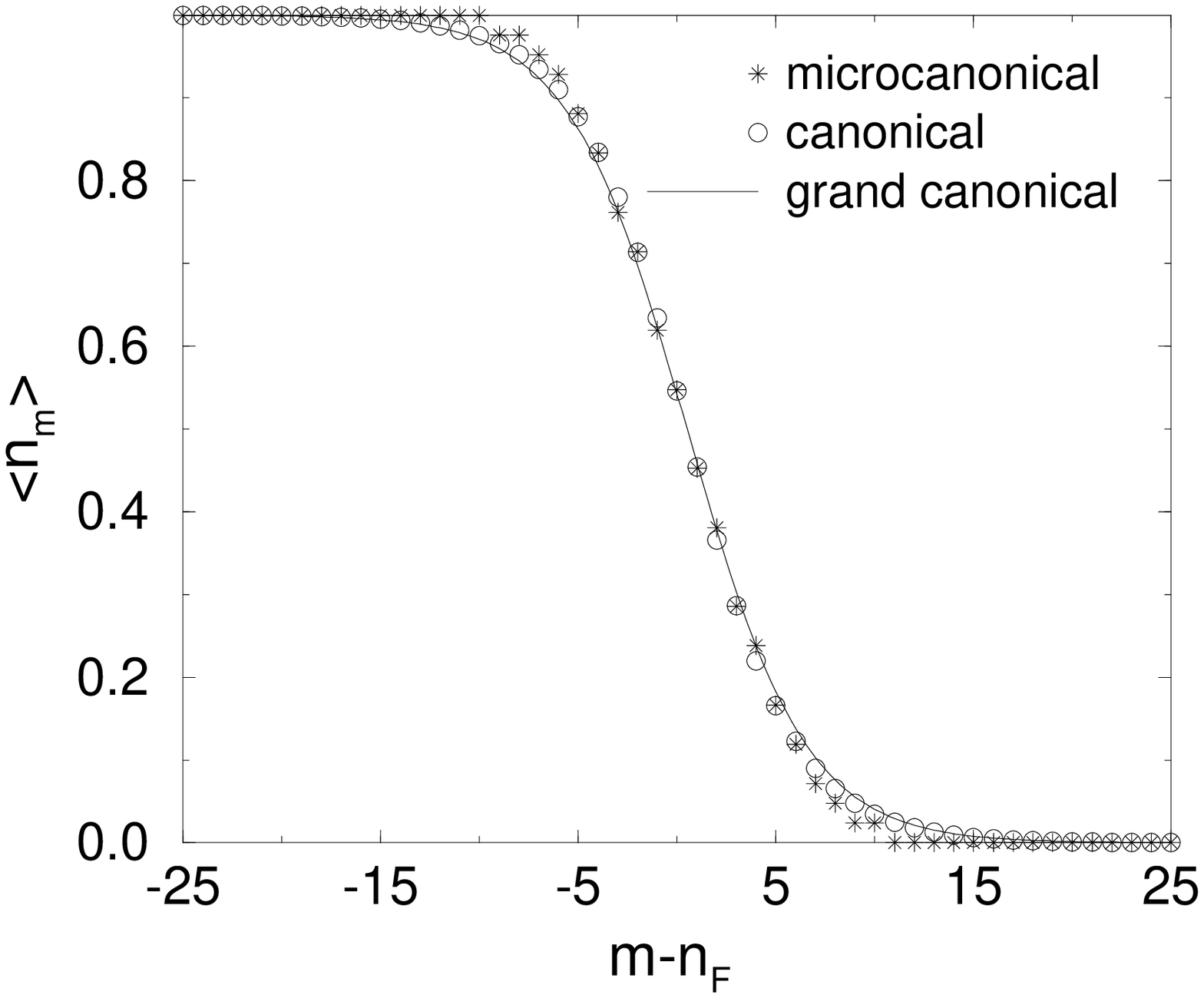}    %Evtl noch \label usw.
\vspace{0.0cm}
\caption{Occupation numbers for the three ensembles for $\tau=3$
respectively $M=10$.}  
\end{figure}   
\begin{figure}[hbt]
\hspace{0.0cm} 
\epsfxsize8.0cm  %Breite des Plots (Siehe auch dvips)
\epsfbox{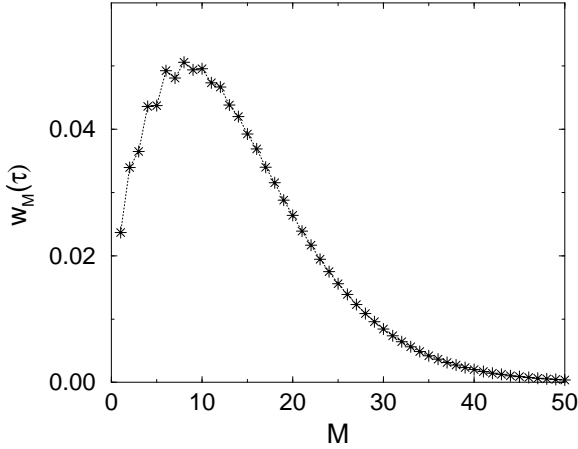}    %Evtl noch \label usw.
\vspace{0.0cm}
\caption{Probability distribution of the energy for $\tau \equiv k_B T/ 
\Delta=3$.}  
\end{figure}

\section{Bosonization of fermionic operators}

This section is more technical than the previous ones and some
familiarity with the method of second quantization is required.
In section II we have shown that the $N$ fermion states of the model
with the linearized dispersion are {\it highly degenerate} and
can be labeled by the list of upward shifts corresponding to 
partitions, which leads to the {\it bosonic} canonical partition
function.
We now show that one can choose {\it linear combinations} of the 
degenerate states for a given $M$, which can be described by
bosonic creation and annihilation operators (ladder operators).
They play the essential role for the solution of the interacting
fermion problem \cite{Tomonaga,Luttinger,ML}. One defines operators
bilinear in the fermionic operators
\begin{equation}\label{33}
b_l^{\dagger}\equiv \frac{1}{\sqrt{l}}\sum_{n=1}^{\infty}
c_{n+l}^{\dagger}c_n ,
\end{equation}
with $l\geq 1$. This is a sum of operators which all shift a fermion
upwards by $l$ energy units.
Using the fermionic anticommutation rules it is easy to show
that $[b_l^{\dagger},b_{l'}^{\dagger}]=0$ and 
\begin{equation}\label{eq:34}
[b_l,b_l^{\dagger}] =\frac{1}{l}\sum_{n=1}^l c_n^{\dagger}c_n          
.
\end{equation}
The operator on the rhs counts the number of fermions in the lowest
$l$ states at the bottom of the band divided by $l$. For $l \ll n_F$
and $N$ fermion states which have {\it no holes at the bottom of the Fermi
see} one can replace the operator on the rhs of Eq.\ (34) by the unit
operator. One can make this
an exact statement if a Dirac sea is added, i.e. the sum in Eq.\ (33)
runs not from $1$ but from $-m_0$ with $m_0 \to \infty$.
An analogous discussion for $l\neq l'$ leads to
\begin{equation}\label{eq:35}
[b_l,b_{l'}^{\dagger}]=\delta_{l,l'} {\sf 1 \hspace{-0.3ex} \rule{0.1ex}{1.6ex}       
     \rule{0.3ex}{0.2ex}} ,
\end{equation}                                            
i.e.\ the operators defined in Eq.\ (33) obey {\it boson commutation}
relations.

With the help of these ladder operators we can create an orthonormal
basis set of many fermion states labeled by {\it bosonic quantum
numbers} $\{\tilde m_l\}$ with $\tilde m_l \in {\rm I \! N}_0$,
if one treats the Fermi sea
$\left\vert F(N)\right>$ as the harmonic oscillator ground state
\begin{equation}\label{eq:36}
\left\vert\{\tilde{m}_l\}\right>=\prod_l\frac{1}{\sqrt{\tilde m_l!}}
(b_l^{\dagger})^{\tilde m_l}\left\vert F(N)\right>   .
\end{equation}
All states with $\sum_l l\tilde m_l=M$
provide a new basis of states corresponding to the excitation
energy $M\Delta$ of $H_0$. They are rather complicated
{\it linear combinations} of the states discussed in section II
which could be labeled either by fermionic occupation numbers or 
by the list $[\alpha]_M$ of the upward shifts. If for $M=4$
we label the $\left\vert [\alpha]_M\right>$
by $\left\vert i\right>$ with $i=1,..,5$, in the
order in which the partitions of $M=4$ were presented in section II, 
we obtain e.g.
\begin{eqnarray}\label{eq:37}
b_4^{\dagger}\left\vert F\right>
& = & \frac{1}{2}\left(\left\vert 1\right>-\left\vert
2\right>+\left\vert 4\right>-\left\vert 5\right>\right)  \nonumber\\
b_3^{\dagger}b_1^{\dagger}\left\vert F\right> & = &
\frac{1}{\sqrt{3}}\left(\left\vert 1\right>-\left\vert 3\right>
+\left\vert 5\right>\right)  \nonumber \\
\frac{1}{\sqrt{4!}}(b_1^{\dagger})^4\left\vert F\right> & = &
\frac{1}{\sqrt{24}}\left(\left\vert 1\right>+3\left\vert 2\right>
\right. \nonumber \\
&& \left. +2\left\vert 3\right>+3\left\vert 4\right>+\left\vert 5\right>
\right).
\end{eqnarray}
As long as we consider only noninteracting fermions both types
of basis sets are completely well suited for the description.
It turns out that for fermions with a two-body interaction 
the interacting groundstate has a much simpler form in
the {\it bosonic} basis set Eq.\ (36).  

In Eq.\ (33) we have expressed the Bose operators $b_l (b_l^{\dagger})$
in terms of the fermionic operators. In order to take advantage of
the bosonization one has to address the reverse problem:
{\it Can one express fermionic operators completely in terms of the
Bose operators}? Here one proceeds in two steps. It is possible 
to {\it bosonize} the operator $H_0$ with the help of the
Kronig identity\cite{Haldane}
\begin{equation}\label{eq:38}
\sum_{l=1}^{\infty}lb_l^{\dagger}b_l=
\sum_{n=1}^{\infty}nc_n^{\dagger}c_n -\frac{1}{2}
(\hat N^2+\hat N) ,
\end{equation}
where $\hat N\equiv \sum_nc_n^{\dagger}c_n
$ is the fermionic particle number operator.
For the proof of this relation the boson commutation relations
of the $b_l(b_l^{\dagger})$ is not needed.
As the lhs of Eq.\ (38) contains products of {\it four} fermion
operators it might look surprising that one obtains a one-particle
operator and only a term involving the particle number operator 
on the right hand side. It is left as an exercise to the reader 
to show that the additional four fermion terms exactly vanish.
If we subtract $n_F\hat N$ and multiply by $\hbar v_F
\pi/L$ we obtain
\begin{equation}\label{39}
H_0=\hbar v_F \sum_{l=1}^{\infty}\frac{l\pi}{L}b_l^{\dagger}b_l
+h(\hat N) ,
\end{equation}
where the function $h(\hat N)$ is irrelevant in the following.

It as also possible to bosonize a long range two-body interaction if
{\it periodic } boundary conditions are used. 
Generalizing the procedure described above one linearizes
around {\it both} Fermi points $\pm k_F$. Then
the Bose operators corresponding to Eq.\ (33) are  
the two contributions to the Fourier
transformed density operators. 
As the two-body interaction is {\it quadratic in the density}
the two-body operator is {\it bilinear} in the Bose operators.
The problem of interacting fermions is therefore reduced to 
a quadratic form in Bose operators, which is easily diagonalized by
a canonical transformation \cite{Tomonaga,ML}. We do not further
discuss interacting fermions here but want to calculate the 
expectation value of $\hat n_m$ in the canonical ensemble
for noninteracting electrons. For
that purpose we have to express $\hat n_m$ in terms of Bose operators.
In order to do that we have to proceed similarly as in the
{\it bosonization of the field operator}\cite{LP,Haldane}.

This requires some preliminary steps. Let us consider the following
operators {\it linear} in the boson operators
\begin{equation}\label{eq:40}
A_+=\sum_{n=1}^{\infty}\alpha_nb_n^{\dagger} \;\; ;
\;\;B_-=\sum_{n=1}^{\infty}\beta_nb_n,
\end{equation}
where $\alpha_n$ and $\beta_n$ are arbitrary constants.  
Using the boson commutation relations Eq.\ (35) one can prove
the commutation relations
\begin{eqnarray}\label{41}
&&[b_m,e^{B_-} e^{A_+}]=\alpha_m e^{B_-} e^{A+}  \nonumber\\
&&[b_m^{\dagger},e^{B_-} e^{A_+}]=-\beta_m e^{B_-} e^{A_+} .
\end{eqnarray}     
If we can construct an operator $S$ in terms of the fermionic operators
which fulfills the commutation relations
$[b_m,S]=- \alpha_m S$ and $[b_m^{\dagger},S]=\beta_m S$
then $S e^{B_-} e^{A_+}$ {\it commutes} with all $b_m$ and
$b_m^{\dag}$ and can therefore only be a function $G_{\hat{N}}$ of the 
number operator $\hat{N}$, which commutes with the boson operators \cite{Schur}, 
i.e.\ 
\begin{eqnarray}
\label{eq:42}
S= G_{\hat{N}} e^{- A_+} e^{-B_-} .
\end{eqnarray}

In order to bosonize $c_n^{\dagger}c_n$ we consider the more
general operator $c_k^{\dagger}c_l$ with $k,l\in 
{\sf Z \hspace{-1ex} Z \hspace{1ex}}$, i.e.\ the
infinite Dirac sea is included,
and determine its commutation
relations with $b_m$ and $b_m^{\dagger}$. 
Using the anticommutation
relations Eq.\ (26) one obtains
\begin{eqnarray}\label{eq:43}
&&[b_m,c_k^{\dagger}c_l]=\frac{1}{\sqrt{m}}(c_{k-m}^{\dagger}c_l
-c_k^
{\dagger}c_{l+m})   \nonumber\\
&&[b_m^{\dagger},c_k^{\dagger}c_l]=\frac{1}{\sqrt{m}}
(c_{k+m}^{\dagger}c_l-c_k^{\dagger}c_{l-m}) .
\end{eqnarray}
If we now multiply Eq.\ (43) by $e^{-iku}e^{ilv}$ with real $u$ and $v$
and sum over $k$ and $l$ the commutation relations with the resulting
operator are of the form as discussed following Eq.\ (41)
\begin{eqnarray}\label{eq:44}
&&[b_m,\psi^{\dagger}(u)\psi(v)]=\frac{1}{\sqrt{m}}
(e^{-imu}-e^{-imv})\psi^{\dagger}(u)
\psi(v) \nonumber\\
&&[b_m^{\dagger},\psi^{\dagger}(u)\psi(v)]= \frac{1}{\sqrt{m}}
(e^{imu}-e^{imv})\psi^{
\dagger}(u)\psi(v) ,
\end{eqnarray}
and the auxiliary field operator $\psi(v)$ is defined as
\begin{equation}\label{eq:45}
\psi(v)=\sum_{l=-\infty}^{\infty}e^{ilv}c_l .
\end{equation}
It is periodic in $v$ with period $2\pi$.
Using the arguments presented following Eq.\ (41) the bosonized form
of $\psi^{\dagger}(u)\psi(v)$ reads
\begin{equation}\label{eq:46}
\psi^{\dagger}(u)\psi(v)=G_{\hat N}(u,v)e^{-i(\phi^{\dagger}
(u)-\phi^{\dagger}(v))}e^{-i(\phi(u)-\phi(v))},
\end{equation}
where the operator $\phi(u)$ is given by
\begin{equation}\label{eq:47}
\phi(u)=-i\sum_{n=1}^{\infty}\frac{e^{inu}}{\sqrt{n}}b_n ,
\end{equation}
and $G_{\hat N}(u,v)$ is a so far undetermined function of the
particle number operator. As the Fermi sea 
with $N$
fermions in addition to the Dirac sea 
is the vacuum state for the bosons, i.e. $b_l\left\vert F(N)\right>
=0$, the expectation value of the product of the 
exponentials in Eq.\ (46) in the Fermi sea equals one and yields
\begin{equation}\label{eq:48}
G_N(u,v)=\left<F(N)\vert \psi^{\dagger}(u)\psi(v)\vert F(N)\right>,
\end{equation}
which can be calculated in the fermionic picture using the
definition Eq.\ (45) and $\left<F(N)\vert c_k^{\dagger}c_l
\vert F(N)\right>=\delta_{k,l}\Theta(N-l)$, where $\Theta(x)$
is the step function ($\Theta(0)=1$)
\begin{eqnarray}\label{49}
G_N(u,v) & = & \sum_{l=-\infty}^N e^{il(u-v)}\nonumber \\
& = & e^{iN(u-v)}\sum_{l=-\infty}
^0 e^{il(u-v-i0)}.
\end{eqnarray}
In the second equality we have added an infinitesimal imaginary
shift in $u-v$ in order to make the formal sum convergent. This is 
a typical example of the problems with the infinite Dirac sea. 

After the bosonization of $\psi^{\dagger}(u)\psi(v)$ is completed,
which actually
is simpler than the bosonization of a single field operator
as no additional operators changing the fermion number have to be
introduced
\cite{Haldane}, we can come to our application of 
calculating the canonical expectation value of $c_m^{\dagger}c_m$.
If we denote the canonical average of an operator $A$
in a state with $N$ fermions
in addition to the Dirac sea by $\left<A\right>_{N,T}$ we obtain
with the Bose function $b(x)\equiv (e^x-1)^{-1}$ 
\begin{eqnarray}\label{eq:50}
&& \left<\psi^{\dagger}(u)\psi(v)\right>_{N,T}  = 
G_N(u,v) \nonumber \\ &&\times \exp{ \left\{ \sum_{n=1}^{\infty}
\left( 2\cos{\left[ n(u-v) \right]}-2\right)
\frac{b(n\Delta\beta)}{n}\right\} }  .
\end{eqnarray}
Here we have used the formula $\left<e^Ae^B\right>=e^{
\frac{1}{2}\left<A^2+2AB+B^2\right>}$ for a canonical
expectation value for free bosons which was proved by Mermin \cite{Mermin} in
``one sentence''. It is valid for operators $A$ and $B$ which are {\it
linear} in
the boson creation and annihilation operators.
To calculate the expectation value of the occupancies we need the
inversion formula to Eq.\ (45)
\begin{equation}\label{eq:51}
c_l=\frac{1}{2\pi}\int_0^{2\pi}e^{-ivl}\psi(v)dv  .
\end{equation}
As the rhs of Eq.\ (50) is a function of $u-v$ only one integration 
can be performed and one obtains
\begin{eqnarray}\label{eq:52}
\left<c_{N+m}^{\dagger}c_{N+m}\right>_{N,T} & = & \frac{1}{2\pi}
\int_0^{2\pi}e^{-iu(m+N)} \nonumber \\ && \times
\left<\psi^{\dagger}(u) \psi(0)\right>_{N,T}.
\end{eqnarray}
In order to analytically perform the remaining integration in Eq.\ (52)
we expand $\left<\psi^{\dagger}(u) \psi(0)\right>$ in a Laurent
series in $z\equiv \exp{(iu)}$
\begin{equation}\label{eq:53}
\left<\psi^{\dagger}(u)\psi(0)\right>_{N,T}=
e^{iuN}\sum_{n=-\infty}^{\infty}d_n e^{iun}        ,
\end{equation}
where the coefficients $d_n$ can be calculated recursively as 
discussed in the appendix\cite{SM}. Using the notation of
section IV we finally obtain
\begin{equation}\label{eq:54}
\left<\hat n_{n_F+m}\right>_{c,T}=d_m .
\end{equation}                  
For all values of $\tau$ for which the calculation of
$\left<\hat n_{m}\right>_{c,T}$  using Eq.\ (31) is numerically 
feasible (e.g.\ $\tau=3$ in Fig.\ 6), the results using Eqs.\ (31) and (54)
are identical. As mentioned in the introduction the result of the 
present section which used Eq.\ (46) as the essential ingredient is {\it very} close
in spirit to the calculation of the occupancies of the {\it interacting} TL-model.
\begin{figure}[hbt]
\hspace{0.0cm} 
\epsfxsize8.0cm  %Breite des Plots (Siehe auch dvips)
\epsfbox{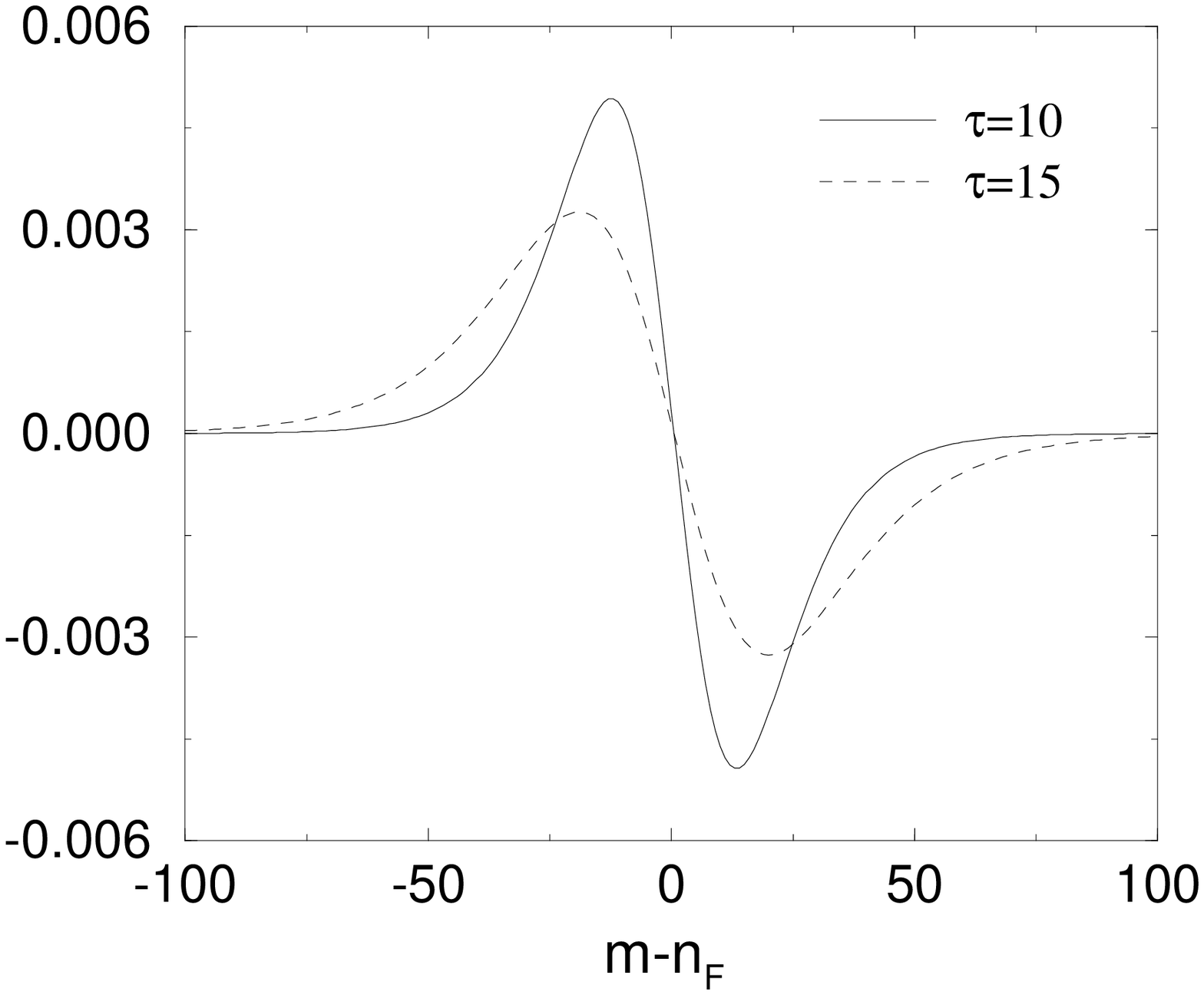}    %Evtl noch \label usw.
\vspace{0.0cm}
\caption{The difference between the canonical occupation numbers and the grand canonical 
Fermi function at $\tau=10$ (solid line) and $\tau=15$ (dashed line).}  
\end{figure}                
Fig.\ 8 shows the difference of the canonical and the grand
canonical result for $\tau =10$ and $15$ as a function of $m-n_F$.
From a plot of the maximum deviation as a function of $1/\tau$ we
find that it scales as $1/\tau$, i.e. $1/L$  if the temperature $T$
is kept constant. 
The relative deviation of the canonical occupation numbers from the Fermi function
$\left| \left[ \left<  n_m \right>_c - f(\epsilon_m) \right] /f(\epsilon_m)
\right|$ for $m>n_F$ increases with $m-n_F$ and saturates in the exponential
tail of the Fermi function. This shows that the canonical occupation numbers decay
as $(1-\eta_{\tau}) \exp{ \{-(m-n_F-0.5)/\tau) \} }$ for $(m-n_F)/\tau \gg 1$.
For $\tau=15$, for example, $\eta_{\tau}=0.032$ and $\eta_{\tau}$ 
goes to zero like $1/\tau$. 
In the thermodynamic limit $L \to \infty$ the sum in the exponent on the rhs of
Eq.\ (50) goes over to an integral which can be performed analytically. Performing
the following integration in Eq.\ (52) one obtains the Fermi function, i.e.\ as 
expected and seen in Fig.\ 8 the canonical occupation numbers approach the grand 
canonical ones for $L \to \infty$. 
 
\section*{Summary}   
This paper intended to serve a dual purpose.
First to explain in terms as simple as possible the {\it concept
of bosonization} which plays the essential role in the Luttinger
liquid picture of interacting fermions in one dimension. We only 
shortly discussed interacting electrons, but every reader who
followed the algebra of section V should have no problem
understanding e.g.
the calculation of the occupancies in the interacting
case\cite{Haldane}. 
As an {\it application of the concepts}  we presented results
for the microcanonical and the canonical ensemble for noninteracting
fermions. The comparison of our results with the standard textbook
grand canonical results can serve as a rare {\it explicit} example
for the general arguments concerning the equivalence of the three ensembles
in the thermodynamic limit.
              
\appendix
\section{}
\label{appendix}
In this appendix we present the method to obtain the coefficients 
$d_m$ of the Laurent series in Eq.\ (53) which determine the canonical
occupation numbers $\left< n_{n_F+m} \right>_{c,T}$. According to
Eqs.\ (49) and (50) we can write 
\begin{eqnarray}\label{eq:a1}
&& \left<\psi^{\dagger}(u)\psi(0)\right>_{N,T}  = 
e^{iNu} \left\{ \sum_{l=0}^{\infty} z^{-l} \right\} \nonumber \\
&& \times \exp{\left\{ f(z)+f(1/z) -2 f(1) \right\}} ,
\end{eqnarray}
where $z=\exp{(iu)}$ and the function $f(z)$ is given by
\begin{eqnarray}
\label{neu1}
f(z)= \sum_{n=1}^{\infty} \frac{b(n/\tau)}{n} z^n .
\end{eqnarray}
The geometric series Eq.\ (A1) is considered as a formal power series.
The first step is to determine the coefficients of the power series of
the function
\begin{eqnarray}
\label{neu2}
F(z) \equiv e^{f(z)} = \sum_{m=0}^{\infty} c_m z^m .
\end{eqnarray}
Using $F'(z)= f'(z) F(z)$ and $c_0=1$ one can obtain 
recursion relations for $m \geq 1$ 
\begin{eqnarray}
\label{neu3}
c_m =  \frac{1}{m} \sum_{l=1}^{m} b(l/\tau)
c_{m-l} .
\end{eqnarray}
The next step is to write the last factor in Eq.\ (A1) as a Laurent
series
\begin{eqnarray}
\label{neu4}
F(z) F(1/z) / \left[F(1) \right]^2 \equiv \sum_{l=-\infty}^{\infty} a_l z^l .
\end{eqnarray}
By comparing coefficients one obtains using Eqs.\ (A2) and (A3) 
\begin{eqnarray}
\label{neu5}
a_l = \exp{\left\{ -2 \sum_{n=0}^{\infty} b(n/\tau)/n   \right\}}
\sum_{m=0}^{\infty} c_m c_{m+l} = a_{-l} .
\end{eqnarray}
As for fixed $\tau$ the $c_m$ go to zero exponentially
for large $m$ the infinite sum converges well. In the last step we
have to multiply the Laurent series in Eq.\ (A5) with the geometric
series in Eq.\ (A1). This yields with Eqs.\ (53) and (54) 
\begin{eqnarray}
\label{neu6}
d_m = \sum_{n=0}^{\infty} a_{n+m} = \left< n_{n_F+m} \right>_{c,T} .
\end{eqnarray}
The numerical evaluation of the 
canonical occupation numbers with this method 
is possible for {\it much} higher temperatures than
via Eq.\ (31).


\begin{thebibliography}{*}

\bibitem{AM} N.\ W.\ Ashcroft and N.\ D.\ Mermin, {\it Solid State Physics}
(Holt, Rinehart and Winston, New York, 1976)
\bibitem{Landau} L.\ D.\ Landau, ``The theory of a Fermi liquid,''
Sov.\ Phys.\ JETP {\bf 3}, 920-925 (1957)
\bibitem{NP}  D.\ Pines and P.\ Nozieres, {\it The Theory of Quantum
liquids} (W.\ A.\ Benjamin, New York, 1966)
\bibitem{Shankar} for a review see R.\ Shankar, ``Renormalization-group approach
to interacting fermions,'' Rev.\ Mod.\ Phys.\ {\bf 66}, 129-192 (1994)
\bibitem{Tomonaga} S.\ Tomonaga, ``Remark on Bloch's method of sound
waves applied to many fermion problems,'' Progr.\ Theor.\ Phys.\ {\bf 5},
544-569 (1950)
\bibitem{reprints} Reprints of most of the important papers in this field
together with an excellent introduction are presented in 
{\it Bosonization,} Editor M.\ Stone (World scientific, Singapore, 1994)
\bibitem{Luttinger} J.\ M.\ Luttinger, ``An exactly soluble model of a 
many-fermion system,'' J.\ Math.\ Phys.\ {\bf 4}, 1154-1162 (1963)
\bibitem{Thirring} W.\ Thirring, ``A soluble relativistic field theory,''
Ann.\ Phys.\ (NY) {\bf 3}, 91-112 (1958)
\bibitem{ML} D.\ C.\ Mattis and E.\ H.\ Lieb, ``Exact solution to a
many-fermion system and its associated boson field,''
J.\ Math.\ Phys.\ {\bf 6}, 304-312 (1965)
\bibitem{LP} A.\ Luther and I.\ Peschel, ``Single particle states, Kohn
anomaly and pairing fluctuations in one dimension,''
Phys.\ Rev.\ {\bf B9}, 2911-2919 (1974)
\bibitem{Haldane} F.\ D.\ Haldane,`` 'Luttinger liquid theory' of one 
dimensional quantum fluids: I.\ Properties of the Luttinger model and
their extension to the general 1D interacting spinless Fermi gas,''
J.\ Phys. C {\bf 14}, 2585-2609 (1981)
\bibitem{Schulz} H.\ J.\ Schulz, ``Correlation exponents and the 
metal-insulator transition in the one-dimensional Hubbard model,''
Phys.\ Rev.\ Lett.\ {\bf 64}, 2831-2834 (1990)   
\bibitem{MS}
V.\ Meden and K.\ Sch\"onhammer, ``Spectral functions for the
Tomonaga-Luttinger model,'' Phys.\ Rev.\ B {\bf 46}, 15753-15760 (1992)
\bibitem{Voit} J.\ Voit, ``Charge-spin separation and spectral
properties of Luttinger liquids,'' Phys.\ Rev.\ B{\bf 47}, 6740-6743 (1993)
\bibitem{Baer} B.\ Dardel, D.\ Malterre, M.\ Grioni, P.\ Weibel, Y.\ Baer, 
J.\ Voit, and D.\ J\'erome, ``Possible observation  of a Luttinger-liquid behaviour 
from photoemission spectroscopy of one-dimensional organic conductors,'' 
Europhys.\ Lett.\ {\bf 24}, 687-692 (1993)
\bibitem{Galic} H.\ Galic, ``Alchemy in 1+1 dimension: From bosons to
fermions,'' Am.\ J.\ Phys.\ {\bf 59}, 1088-1096 (1991)
\bibitem{Reif} e.g.\ F.\ Reif, {\it Fundamentals of Statistical and
Thermal Physics} (MacGraw-Hill, New York, 1965)
\bibitem{terHaar}D.\ ter Haar, {\it Elements of Statistical Mechanics}
(Rinehart, New York 1954)
\bibitem{Euler} L.\ Euler, ``De partitione numerorum,'' (1753) in
{\it Leonardi Euleri opera omnia} (Teubner, Leipzig, 1911)
\bibitem{Andrews} G.\ E.\ Andrews, The Theory of Partitions
(Addison-Wesley, Reading, 1976)  
\bibitem{SM} Partitions were used in the context of calculating
spectral functions in the interacting system in:
K.\ Sch\"onhammer and V.\ Meden, ``Nonuniversal spectral
properties of the Luttinger model,'' Phys.\ Rev.\ B {\bf 47}, 16205-16215 (1993)
\bibitem{Abramowitz} M.\ Abramowitz and I.\ A.\ Stegun, {\it Handbook of
Mathematical Functions} (Dover, New York, 1970)   
\bibitem{Schur} A similar argument is well known in group theory 
and is called ``Schur's Lemma'', see e.g.\ M.\ Tinkham, {\it Group Theory and 
Quantum Mechanics} (MacGraw-Hill, New York, 1964)
\bibitem{Mermin} N.\ D.\ Mermin, ``A short simple evaluation of
expressions of the Debye-Waller form,''  J.\ Math.\ Phys.\ {\bf 7}, 1038
(1966)
\end{thebibliography}
\end{document}